\documentclass[aps,pra,showpacs,amssymb,twocolumn]{revtex4-1}
\usepackage{graphicx}
\usepackage{appendix}
\usepackage{color}
\usepackage[caption=false]{subfig}
\usepackage{epstopdf} 
\DeclareGraphicsExtensions{.pdf,.eps,.png,.jpg,.mps}
\usepackage[colorlinks=true,citecolor=blue,urlcolor=blue,linkcolor=blue]{hyperref}
\usepackage{bm,amsmath}
\usepackage{dcolumn}

\newcommand{\bra}[1]{\left\langle #1 \right|}
\newcommand{\ket}[1]{\left|#1\right\rangle}

\def\BEq{\begin{equation}}
\def\EEq{\end{equation}}
\def\BEqA{\begin{eqnarray}}
\def\EEqA{\end{eqnarray}}
\def\BW{\begin{widetext}}
\def\EW{\end{widetext}}

\begin{document}

\title{Simulating Anderson localization via quantum walk on a one-dimensional lattice of superconducting qubits}

\author{Joydip Ghosh}
\email{ghoshj@ucalgary.ca}
\affiliation{Institute for Quantum Science and Technology, University of Calgary, Calgary, Alberta T2N 1N4, Canada}

\date{\today}

\begin{abstract}

Quantum walk (QW) on a disordered lattice leads to a multitude of interesting phenomena, such as Anderson localization. While QW has been realized in various optical and atomic systems, its implementation with superconducting qubits still remains pending. The major challenge in simulating QW with superconducting qubits emerges from the fact that on-chip superconducting qubits cannot hop between two adjacent lattice sites. Here we overcome this barrier and develop a novel gate-based scheme to realize the discrete time QW by placing a pair of qubits on each site of a 1D lattice and treating an excitation as a walker. It is also shown that various lattice disorders can be introduced and fully controlled by tuning the qubit parameters in our quantum walk circuit. We observe a distinct signature of transition from the ballistic regime to a localized QW with an increasing strength of disorder. Finally, an eight-qubit experiment is proposed where the signatures of such localized and delocalized regimes can be detected with existing superconducting technology. Our proposal opens up the possibility to explore various quantum transport processes with promising superconducting qubits. 

\end{abstract}


\maketitle

\section{Introduction}
\label{sec:Introduction}

Quantum Walk (QW) was first proposed by Aharonov et al. \cite{PhysRevA.48.1687} and has since then remained a subject of growing interest for many subsequent theoretical \cite{1367-2630-10-5-053025,Meyer1996JSP,Daharonov2001STOC,Childs2002QIP,PhysRevA.65.032310,0305-4470-35-12-304,PhysRevA.67.042305,PhysRevA.58.915,PhysRevLett.102.180501,Inui2005IJQI,PhysRevLett.93.180601,PhysRevA.70.022314,PhysRevA.78.042334,PhysRevA.67.052307,Childs15022013} and experimental \cite{PhysRevA.67.042316,PhysRevA.72.062317,Karski10072009,PhysRevLett.106.180403,PhysRevLett.103.090504,PhysRevLett.104.100503,PhysRevLett.104.050502,Crespi2013,Jeong2013} pursuits. QW has already turned out to be significant in developing quantum algorithms with polynomial as well as exponential speedups (See Ref.~\cite{doi:10.1142/S0219749903000383} for a review), designing a universal model for quantum computation \cite{PhysRevLett.102.180501,Childs15022013}, and studying various quantum transport processes \cite{Schwartz2007,PhysRevLett.93.180601,1367-2630-11-3-033003,Mulken201137}. While QW has so far been demonstrated with NMR \cite{PhysRevA.67.042316,PhysRevA.72.062317}, neutral atoms \cite{Karski10072009}, trapped ions \cite{1367-2630-12-12-123016,PhysRevLett.103.090504,PhysRevLett.104.100503}, and optical systems \cite{PhysRevLett.110.263602,PhysRevLett.104.050502,PhysRevLett.106.180403,Crespi2013,Jeong2013}, its realization with superconducting qubits still remains pending \footnote{An implementation scheme for QW with superconducting circuit QED systems is, however, proposed in Ref.~\cite{1367-2630-10-5-053025} and \cite{PhysRevA.78.042334}.}. Superconducting qubits are composed of on-chip Josephson junctions and therefore, unlike many other qubit realizations, cannot hop from one lattice site to another that remained a major challenge to implement QW with such systems.

\begin{figure}[htb]
\centering
\subfloat[]{\includegraphics[angle=0,width=\linewidth]{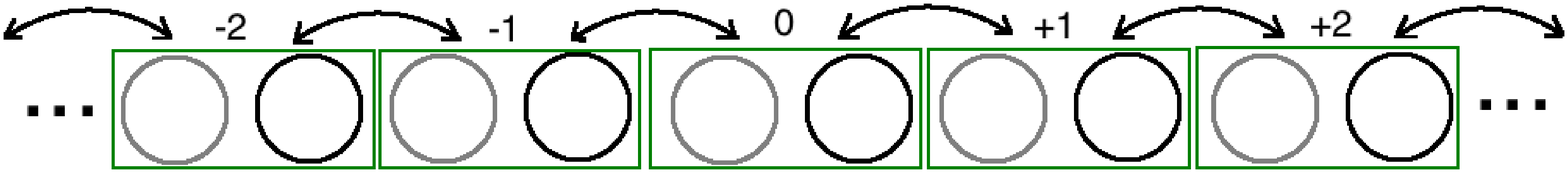}} \\
\subfloat[]{\includegraphics[angle=0,width=\linewidth]{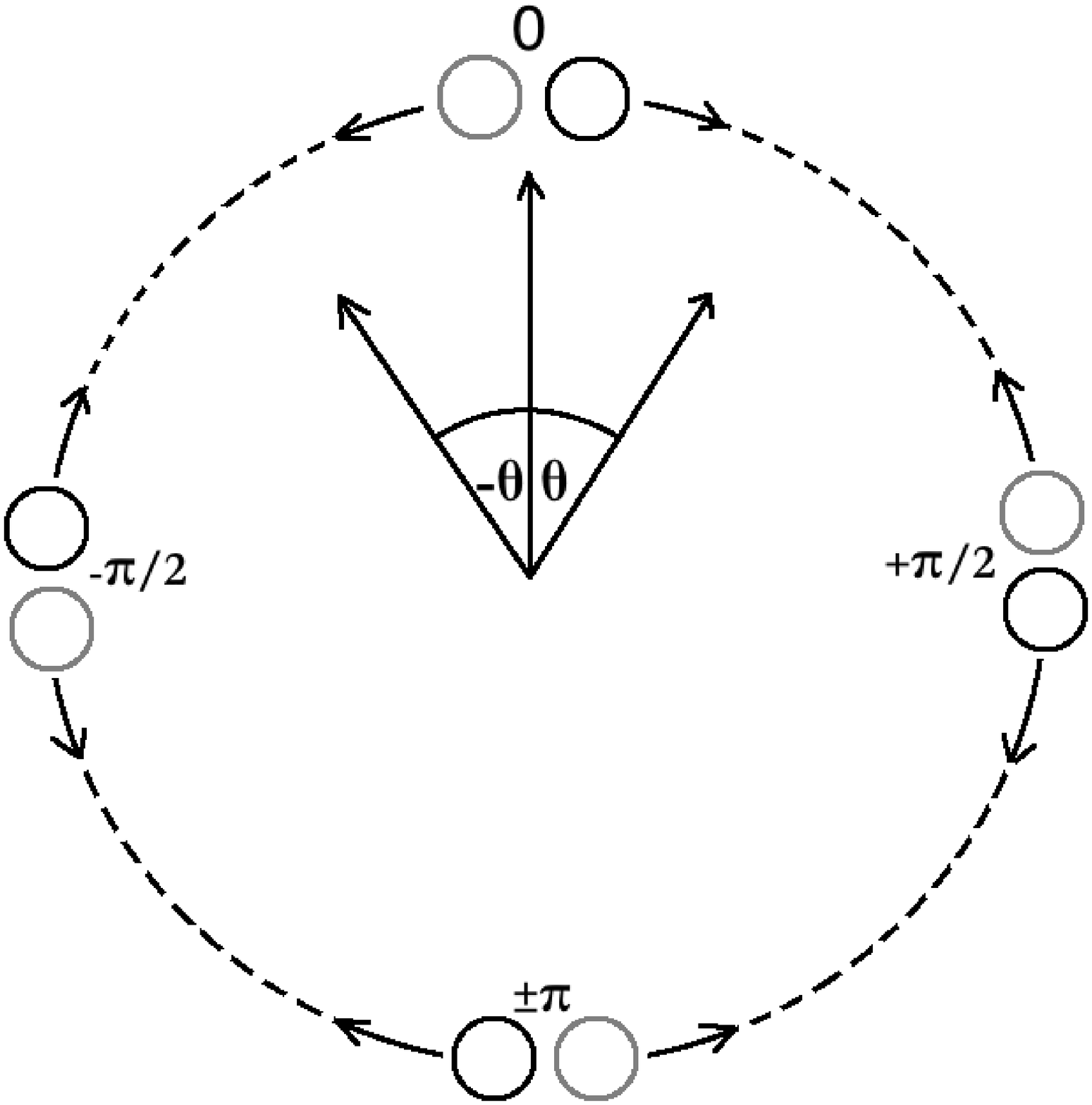}}
\caption{(Color online) A schematic diagram describing QW (a) on a line, and (b) on a circle. In both cases, every single lattice site consists of two (gray and black) superconducting qubits. The arrows denote possible hopping to the nearest-neighbor. In (a) each lattice site is characterized with an integer (positive on the right and negative on the left from the origin) and in (b) each site is characterized by the angle from the origin (positive along clockwise and negative along anticlockwise direction).}
\label{fig:QRWSC}
\end{figure}

\begin{figure*}[htb]
\includegraphics[angle=0,width=\linewidth]{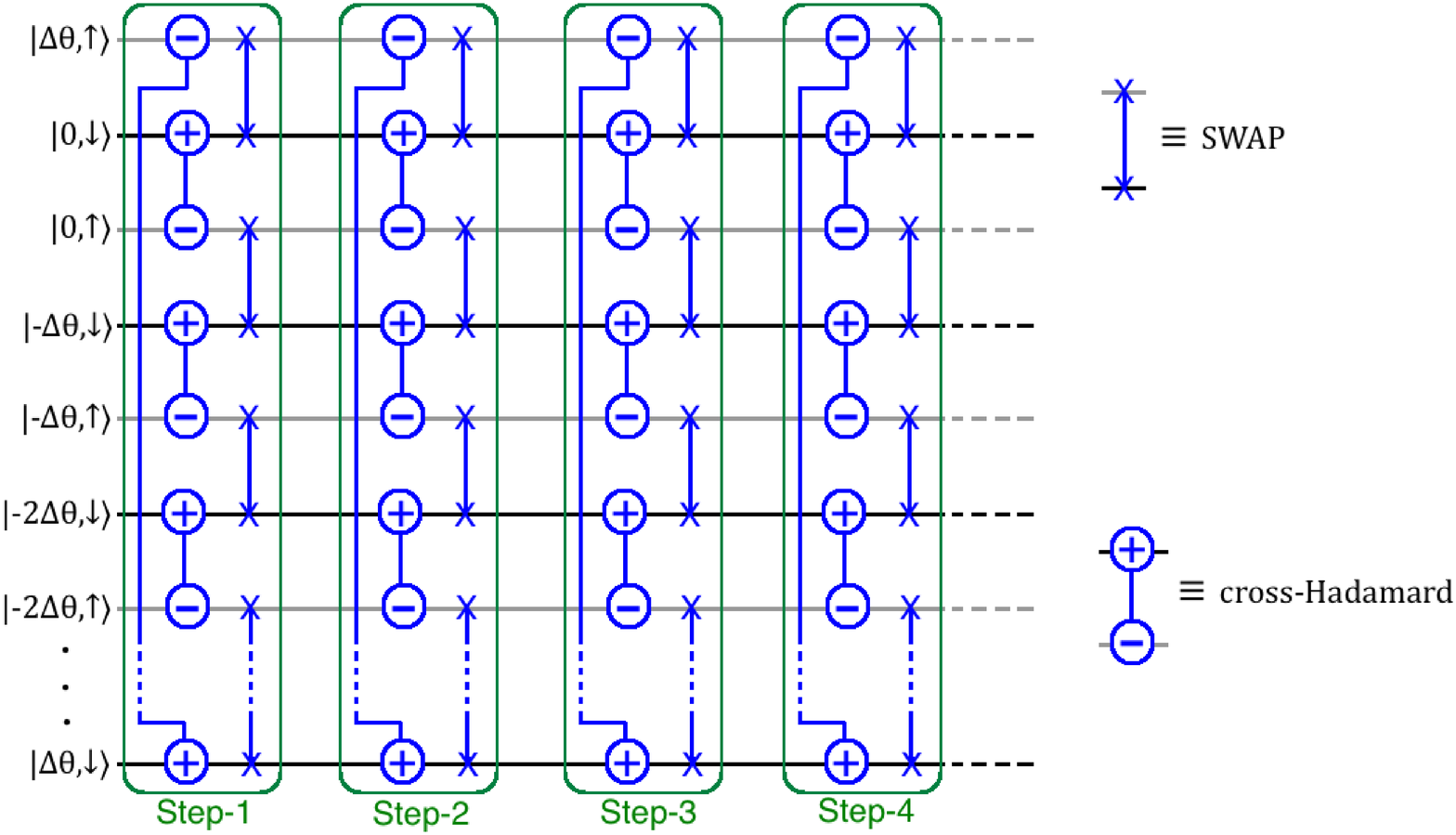}
\caption{(Color online) The quantum circuit diagram for the DTQW on a circular lattice. The horizontal axis shows consecutive timesteps and the vertical axis shows qubits on the lattice with $+\Delta\theta$ ($-\Delta\theta$) being angular separation between two adjacent lattice sites along clockwise (anticlockwise) direction.}
\label{fig:generalCircuit}
\end{figure*}

We here propose a gate-based approach to simulate Discrete Time QW (DTQW) that circumvents the existing challenges. Our approach is applicable to any qubit realization (capable to demonstrate single- and two-qubit gates) and specifically useful for stationary qubits (qubits that cannot hop between lattice sites, such as semiconductor spin qubits or superconducting qubits). We, however, concentrate on superconducting qubits here primarily due to their high degree of scalability (required to fabricate long lattices) and long coherence times (required to simulate long transport processes). In this section, we outline the overview of our protocol (see Sec.~\ref{sec:QRW} for a brief review on discrete quantum walk). Fig.~\ref{fig:QRWSC} depicts the key idea of our scheme for both the linear and circular lattices. Every empty circle denotes a qubit and arrows between two adjacent sites indicate possible nearest-neighbor hopping of the excitation at any given time. We treat the excitation itself as a walker and define it to be `spin-up' (denoted by $\uparrow$) if it is trapped in a gray qubit, and `spin-down' (denoted by $\downarrow$) if in a black qubit. Since the excitations of superconducting qubits possess no internal \emph{spin degrees of freedom}, such an arrangement is sufficient to construct the two-dimensional Hilbert space for the so-called `coin-tossing' operation. For DTQW, these coin tossing operations (usually Hadamard) determine the next hopping direction (left or right for the linear lattice and clockwise or anticlockwise for the circular one) of the walker at each site.

For the purpose of this work, we concentrate on a single-particle DTQW on a circle, while our protocol can be extended for a linear array simply by imposing an open boundary condition at any site on the circular lattice. The required architecture for our scheme consists of a circular array of superconducting transmon qubits with nearest-neighbor couplings via tunable couplers \cite{Hime01122006,Niskanen04052007,PhysRevLett.104.177004,PhysRevLett.106.060501}. Fig.~\ref{fig:generalCircuit} shows the circuit diagram for the quantum walk, where the horizontal direction denotes time, and the vertical direction denotes the sites on a circular lattice (with first and last qubits being nearest neighbors). Note that each quantum gate is performed between adjacent qubits in a circular geometry. The state $\ket{\theta,s} (s \in \{\uparrow,\downarrow\})$ denotes the angular position and \emph{effective spin} ($\uparrow$ if the excitation is in a gray qubit, and $\downarrow$ if in a black qubit) of the excitation. $\Delta\theta$ is the angular separation (assumed to be uniform) between neighboring sites. Hopping operations are performed with simultaneous {\sf SWAP} gates between each neighboring lattice sites, and the Hadamard coin-tossing operations are performed with {\sf Hadamard} gates defined on the single excitation subspace (hereafter referred to as {\sf cross-Hadamard} gates) of the pair of qubits at each lattice site. In the two-qubit computational basis ($\{\ket{00},\ket{01},\ket{10},\ket{11}\}$), the {\sf SWAP} gate is defined as,
\BEq
{\sf SWAP} \equiv \left[ \begin{array}{cccc}
1 & 0 & 0 & 0 \\
0 & 0 & 1 & 0 \\
0 & 1 & 0 & 0 \\
0 & 0 & 0 & 1 \end{array} \right].
\label{eq:SWAP}
\EEq
{\sf SWAP} is a symmetric operation under the exchange of qubit indices and therefore, it is not required to distinguish between the two qubits. However, an additional care is required in defining the {\sf cross-Hadamard} operation, primarily due to two reasons: First, {\sf cross-Hadamard} is not a symmetric operation under the exchange of qubit indices and therefore, we need to use an explicit notation to distinguish between participating qubits. Second, in our scheme every hopping operation also flips the spin-state of the walker (because in our arrangement every black qubit is coupled to a gray one and vice-versa) and therefore, in order to be consistent with the existing convention, our definition of {\sf cross-Hadamard} gate must take this fact into account and nullify it with an additional internal {\sf SWAP} gate. In order to comply with these constraints we adopt the notation for {\sf cross-Hadamard} as shown in Fig.~\ref{fig:generalCircuit}, and in the basis
\BEq
\{\ket{00},\ket{01},\ket{10},\ket{11}\}, \nonumber
\EEq
the {\sf cross-Hadamard} gate is defined as,
\BEq
{\sf cross\mbox{-}Hadamard} \equiv \left[ \begin{array}{cccc}
1 & 0 & 0 & 0 \\
0 & \frac{1}{\sqrt{2}} & \frac{1}{\sqrt{2}} & 0 \\
0 & -\frac{1}{\sqrt{2}} & \frac{1}{\sqrt{2}} & 0 \\
0 & 0 & 0 & 1 \end{array} \right].
\label{eq:cross-Hadamard}
\EEq
In our scheme, the excitation itself, therefore, plays the role of a walker and, under the proposed gate operations, performs a DTQW via the constructive and destructive interferences of various paths.

It is also possible to introduce disorders on the lattice sites within our scheme. QW in presence of static and random lattice disorders results in a localization of the wavefunction of the walker. For an infinitely long lattice, the existence of such localized eigenmodes was predicted by Anderson in 1958 \cite{PhysRev.109.1492}, which eventually turned out to be an ubiquitous effect in any form of energy transport through disordered lattices. Since its discovery, Anderson localization has so far been reported for light waves \cite{Wiersma1997,Schwartz2007,PhysRevLett.100.013906}, microwaves \cite{Dalichaouch1991,Chabanov2000}, acoustic waves \cite{Weaver1990129}, and matter waves \cite{Billy2008}. It still remains a topic of ongoing discussions if the terminology--Anderson localization--should be used for localized quantum walks on a finite disordered lattice where the initial wavefunction of the walker is also localized \cite{Konno2010QIP,Wojcik2012PRA,PhysRevA.76.012315,PhysRevA.87.012314,Konno2013QIP,joye2010JSP}. We demonstrate here that if the walker is prepared initially on a particular lattice site, then it gets localized around its initial location if we perform QW in presence of a random static disorder on each site. We also observe that if the walker is prepared initially on two diametrically opposite lattice sites (in uniform superposition as shown in Fig.~\ref{fig:doublelocalization32sites50runs}), the final wavefunction of the walker gets localized around the initially populated sites when disorder is turned on. We, however, refer to this effect as Anderson localization in this work (following Ref.~\cite{Crespi2013} and \cite{PhysRevLett.106.180403}). While we can introduce both static and dynamic disorders using this approach, we here mainly focus on static disorders and discuss how the signature of wavefunction localization can be extracted for a finite-sized 1D lattice. The random static diagonal disorders are introduced in our quantum walk by inserting a $\sigma^{z}$ rotation on each qubit after every {\sf cross-Hadamard} operation. The $\sigma^{z}$ rotation angles are time-independent but chosen from a uniform random distribution between $-W\pi$ and $W\pi$ ($W$ is referred to as the \emph{disorder strength} and assumed to vary between $0$ and $1$) for each lattice site. For superconducting qubits, as described later, these rotations are in fact performed by random excursions of qubit frequencies, which is similar to assigning a random energy on each lattice site in Anderson's original tight-binding model \cite{PhysRev.109.1492}.

The rest of the paper is organized as follows. In Sec.~\ref{sec:QRW}, we provide a detailed review of DTQW on a 1D lattice. The connection between quantum walk and localization is also discussed in this context. In Sec.~\ref{sec:QRWscq}, we elaborate our scheme to implement DTQW on a circular lattice of superconducting qubits with or without disorders, and numerical results are obtained by simulating the quantum circuit \ref{fig:generalCircuit} for various disorder strengths. The protocols to design the required gate operations are also discussed and an $8$-qubit experiment is proposed to observe localization that is within reach of current superconducting technology. We conclude in Sec.~\ref{sec:conclusions} with possible future directions of this research.

\section{Discrete time quantum walk}
\label{sec:QRW}
In this section, we review the standard protocol for DTQW on a one-dimensional lattice. While in this work we mostly concentrate on the quantum walk on a circle, we still discuss the QW on a linear lattice for completeness.

\subsection{DTQW on a line}

\begin{figure}[htb]
\includegraphics[angle=0,width=\linewidth]{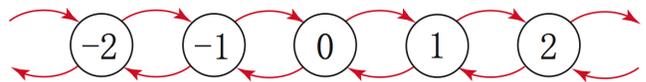}
\caption{(Color online) A schematic diagram showing DTQW on a line. The integers denote the site indices and the arrows denote possible hopping between adjacent sites.}
\label{fig:QRWline}
\end{figure}

In this section, we briefly describe the quantum walk on a line. A detailed discussion on QW on a line can be found in Ref.~\cite{doi:10.1080/00107151031000110776}. Fig.~\ref{fig:QRWline} shows a schematic diagram for DTQW on a line. Each empty circle denotes a lattice site labelled by an integer (varying between $-n$ and $n$, where $2n+1$ is the total number of sites on the lattice) denoting its position. The arrows denote possible nearest-neighbor hopping. We assume that the walker has its own spin degree of freedom and it could either be in $\ket{k,\uparrow}$ or in $\ket{k,\downarrow}$ state, where $k \in \{-n, -(n-1), \ldots, -1, 0, 1, \ldots, n-1, n\}$. In order to have a quantum walk, we assume that at $t=0$ the walker is in $\ket{0,\downarrow}$ state, and perform the Hadamard coin-tossing operation (denoted by $\hat{H}$) on the spin space, which is defined as,
\BEqA
&& \hat{H}\ket{k,\downarrow}=\frac{\ket{k,\downarrow}+\ket{k,\uparrow}}{\sqrt{2}} \nonumber \\
&& \hat{H}\ket{k,\uparrow}=\frac{\ket{k,\downarrow}-\ket{k,\uparrow}}{\sqrt{2}}
\label{eq:coinToss}
\EEqA
For an initial $\ket{\uparrow}$ or $\ket{\downarrow}$ state, this operation creates a uniform superposition. In order to retrieve a classical random walk, the spin of the walker gets measured after this coin-tossing and a shift to the left nearest-neighbor site is performed if the measured state is $\ket{\uparrow}$, and a rightward shift is performed otherwise. The real difference between classical and quantum walk emerges from the fact that the spin of the walker never gets measured for quantum case. Instead, for quantum walk we perform a conditional shift operation (denoted by $\hat{S}$) defined as follows,
\BEqA
&& \hat{S}\ket{k,\downarrow}=\ket{k+1,\downarrow} \nonumber \\
&& \hat{S}\ket{k,\uparrow}=\ket{k-1,\uparrow}
\label{eq:lineShift}
\EEqA
A single \emph{step} in DTQW consists of a Hadamard coin-tossing operation followed by the conditional shift. Table \ref{table:QRWline} shows the probability distribution for a quantum walk on a line after each step. The initial state is assumed to be $\ket{0,\downarrow}$. While for classical random walk the nonzero terms in the probability distribution at every step can be obtained from Pascal's triangle and, therefore, symmetric about the origin, for DTQW we observe a clear departure from the classical case starting from the $3^{\rm rd}$ step. This is a typical characteristic of QW originating from the quantum interference among various possible paths. Another quintessential signature of quantum walk is the standard deviation of the probability distribution at each timestep that scales linearly with the time, as opposed to classical random walk where the variance scales linearly with time instead of standard deviation as,
\BEqA
\sigma_{\rm quantum} &\sim& {\rm total \; timesteps}, \nonumber \\
\sigma_{\rm classical} &\sim& \sqrt{\rm total \; timesteps},
\EEqA
where $\sigma$ denotes the standard deviation. It has been extensively verified numerically that $\sigma_{\rm quantum}$ not only scales linearly with the number of timesteps, but also almost independent of the initial state of the walker \cite{PhysRevA.65.032310}. The linear scaling of standard deviation for quantum walk denotes the \emph{ballistic spread} of the probability distribution of the walker in comparison to its classical diffusion. 

\begin{center}
\begin{table}[htb]
\caption{The probability distribution of a DTQW on a line for various timesteps. A departure from that of classical random walk is observed from the $3^{\rm rd}$ step.\newline}
\resizebox{0.55\linewidth}{!}{
 \begin{tabular}{| c || c | c | c | c | c | c | c |}
  \cline{1-8}
Time & \multicolumn{7}{c |}{Lattice sites} \\
    \cline{2-8}
    steps & -3 & -2 & -1 & 0 & 1 & 2 & 3 \\ \hline
    $0$ & 0 & 0 & 0 & 1 & 0 & 0 & 0 \\ \hline
    $1$ & 0 & 0 & $\frac{1}{2}$ & 0 & $\frac{1}{2}$ & 0 & 0 \\ \hline
    $2$ & 0 & $\frac{1}{4}$ & 0 & $\frac{1}{2}$ & 0 & $\frac{1}{4}$ & 0 \\ \hline
    $3$ & $\frac{1}{8}$ & 0 & $\frac{5}{8}$ & 0 & $\frac{1}{8}$ & 0 & $\frac{1}{8}$ \\ \hline
  \end{tabular}}
\label{table:QRWline}
\end{table}
\end{center}

\subsection{DTQW on a circle}
\label{sec:DTQRWonCircle}

\begin{figure}[htb]
\includegraphics[angle=0,width=\linewidth]{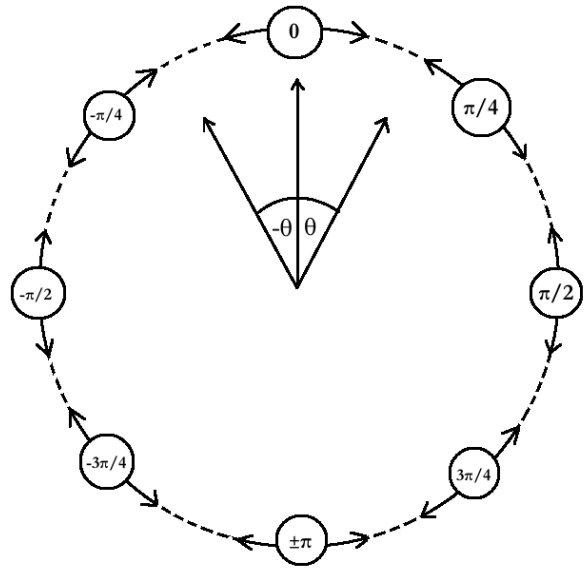}
\caption{A schematic diagram showing DTQW on a circle. The arrows denote possible hopping between adjacent sites and each site is characterized by the angular distance from the origin (positive along clockwise and negative along anticlockwise direction).}
\label{fig:QRWcircle}
\end{figure}

Fig.~\ref{fig:QRWcircle} shows the quantum walk on a circular lattice. Like the linear case, each open circle here denotes a lattice site. We assume the site at the north pole as our origin and identify each site with its angular distance $\theta$ ($\theta \in [-\pi,+\pi]$) from the origin. The Hadamard coin-tossing operation is defined as given by Eq.(\ref{eq:coinToss}), where $k$ denotes the site index on the circular lattice for this case. The conditional hopping is defined as,
\BEqA
&& \hat{S}\ket{\theta,\downarrow}=\ket{\theta+\Delta\theta,\downarrow} \nonumber \\
&& \hat{S}\ket{\theta,\uparrow}=\ket{\theta-\Delta\theta,\uparrow},
\label{eq:circleShift}
\EEqA
with $\Delta\theta$ being the angular separation between two neighboring sites. The probability distribution for circular case remains identical to the linear case until the population hits the boundary, which is at an angle $\pm\pi$ in Fig.~\ref{fig:QRWcircle}. Therefore, prior to the timestep when population gets closer to the boundary from both directions, the standard deviation of the probability distribution at each timestep scales linearly with time as in the case of QW on a line. We demonstrate this signature with our superconducting circuit in Sec.~\ref{sec:QRWscqNoDisorder}. If we start with an initial state $\ket{0,\downarrow}$, then for $|\Delta\theta|=\pi/2$, we discover the walker at $\theta=\pi/2$ after two steps with unit probability due to constructive quantum interference. Again, this is a distinct feature of quantum walk on a circle.

In order to investigate the DTQW analytically on a circle, let us now introduce the so-called Transfer-matrix approach, which especially turns out to be useful later in understanding the localized eigenstates in presence of random disorder. A Transfer-matrix is the matrix representation of an operator that transforms the wavefunction at $j^{\rm th}$ timestep to the wavefunction at $(j+1)^{\rm th}$ timestep. Let $\Psi_{\rm N}(j)$ be the wavefunction of a particle performing quantum walk at $t=j$ on a circular lattice having $N$ sites. The action of Transfer-matrix $T_{\rm N}$ on the $j^{\rm th}$ state can be defined as,
\BEq
\Psi_{\rm N}(j+1)=T_{\rm N}\Psi_{\rm N}(j).
\EEq
In the basis (hereafter referred to as \emph{clockwise} basis), 
\BEq
\{\ket{0,\downarrow},\ket{0,\uparrow},\ket{\Delta\theta,\downarrow},\ket{\Delta\theta,\uparrow}, \ldots, \ket{-\Delta\theta,\downarrow},\ket{-\Delta\theta,\uparrow}\}, \nonumber
\EEq
the matrix representation of $T_{\rm N}$ is given by,
\BEq
T_{\rm N} \equiv \left[ \begin{array}{ccccccc}
0 & A & 0 & \ldots & \ldots & 0 & B \\
B & 0 & A & 0 & \ldots & \ldots & 0 \\
0 & B & 0 & A & 0 & \ldots & 0 \\
\vdots & \vdots & \vdots & \vdots & \vdots & \vdots & \vdots \\
0 & \ldots & 0 & B & 0 & A & 0 \\
0 & \ldots & \ldots & 0 & B & 0 & A \\
A & 0 & \ldots & \ldots & 0 & B & 0 \end{array} \right],
\EEq
where
\BEqA
A &\equiv& \frac{1}{\sqrt{2}}\left[\begin{array}{cc}
0 & 0 \\
1 & -1 \end{array}\right], \nonumber \\ \nonumber \\
B &\equiv& \frac{1}{\sqrt{2}}\left[\begin{array}{cc}
1 & 1 \\
0 & 0 \end{array}\right], \nonumber \\ \nonumber \\
0 &\equiv& \left[\begin{array}{cc}
0 & 0 \\
0 & 0 \end{array}\right].
\EEqA
Note that, $T_{\rm N}$ is a $2N\times2N$ dimensional time-independent matrix and by repeated application of the Transfer-matrix on the initial state, we can express the state of the walker at $k^{\rm th}$ timestep as,
\BEq
\Psi_{\rm N}(k)=(T_{\rm N})^{k}\Psi_{\rm N}(0),
\EEq
$\Psi_{\rm N}(0)$ being the initial wavefunction of the walker. Aharonov et al. \cite{Daharonov2001STOC} and Bednarska et al. \cite{Bednarska200321} showed how to obtain the eigenvalues and eigenvectors of the Transfer-matrix $T_{\rm N}$ analytically, which in fact enables one to determine the $\Psi_{\rm N}(k)$ analytically (at least in principle) for a given initial state. These formulae, however, are not relevant for our purpose and we do not attempt to review these results here.

An interesting point to note in this context is the \emph{recurrence} of a cycle. This is a typical feature of DTQW on a circle or any closed graph topologically equivalent to a circle. For $\Delta\theta=\pi/2$ ({\it i.e.}, when we have $4$ lattice sites on a circle) it is easy to check that the probability distribution repeats itself after every $7$ steps, and therefore, if we start with our walker at the origin initially, the probability of observing it at the origin again becomes unity after every $7$ timesteps. Such a recurrence of probability distribution can be observed after every $23$ timesteps for $\Delta\theta=\pi/4$ (In this work we always assume $\Delta\theta$ to be in the form of $\pi/2^{l}$, $l$ being an integer). However, for $l > 2$ (equivalently, if we have more than $8$ lattice sites on the circle), such a complete recurrence does not occur, while we can always retrieve our walker in the origin with some fractional probability, a phenomenon sometimes referred to as \emph{fractional recurrence} [See Ref.~\cite{chandrasekhar2010} for a detailed analysis of recurrence].

\subsection {Localization of wavefunction in 1D disordered lattice: A perspective from quantum walk}
\label{sec:DTQRWwithDisorder}

In this section, we first consider an infinitely long 1D lattice, investigate the localized eigenmodes in presence of random disorders, and then discuss how to generate such localized eigenstates via quantum walk in a disordered lattice with finite number of lattice sites. 

\subsubsection{Tight-binding model}
Let us consider the the tight-binding model on a 1D lattice for which the Hamiltonian is given by,
\BEq
H_{\rm TB}=\sum_{j}\epsilon_{j}\ket{j}\bra{j}+\sum_{{\langle}jk\rangle}V_{jk}\ket{j}\bra{k},
\label{eq:TBHamiltonian}
\EEq
with $\epsilon_{j}$ being on-site energies, $V_{jk}$ being the coupling between $j^{\rm th}$ and $k^{\rm th}$ sites, and $\langle\ldots\rangle$ denotes nearest-neighbor. For any pair of sites, we here assume,
\BEq
V_{jk} = \left\{ \begin{array}{rl}
  V &\mbox{ if $|j-k|=1$} \\
  0 &\mbox{ otherwise.}
       \end{array} \right.
\EEq
Let $\Psi_{\rm TB}$ be an eigenstate of our tight-binding Hamiltonian (\ref{eq:TBHamiltonian}). In the usual lattice basis, we can express,
\BEq
\ket{\Psi_{\rm TB}}=\sum_{j=-\infty}^{\infty}\psi_{j}\ket{j}.
\EEq
Note that, the probability amplitudes $\psi_{j}$ are real, as our Hamiltonian $H_{\rm TB}$ is assumed to be real and symmetric. The off-diagonal coupling terms $V_{jk}$ (assumed to take a constant value $V$ for our case) contribute to hopping from one lattice site to its neighboring sites and the random diagonal disorders can be introduced in the Hamiltonian by choosing the values of $\epsilon_{j}$ from a uniformly distributed random numbers. The eigenvalue equation for Hamiltonian (\ref{eq:TBHamiltonian}) can be written as,
\BEq
H_{\rm TB}\ket{\Psi_{\rm TB}}=E_{\rm TB}\ket{\Psi_{\rm TB}},
\label{eq:HTBeigen}
\EEq
with $E_{\rm TB}$ being the eigenvalue of our tight-binding Hamiltonian with eigenstate $\ket{\Psi_{\rm TB}}$. 

\subsubsection{Localized eigenstates: Random Matrix Theory}
With a little algebra, we can rewrite Eq.(\ref{eq:HTBeigen}) in the form of recurrence relations of the probability amplitudes as,
\BEq
\epsilon_{j}\psi_{j}+V(\psi_{j-1}+\psi_{j+1})=E_{\rm TB}\psi_{j},
\EEq
for all $j$ on the lattice. In matrix form,
\BEq
\left(\begin{array}{c}
\psi_{j+1} \\
\psi_{j} \end{array} \right)=T_{\rm TB}^{(j)}\left(\begin{array}{c}
\psi_{j} \\
\psi_{j-1} \end{array} \right),
\EEq
where the transfer matrices $T_{\rm TB}^{(j)}$ are defined as,
\BEq
T_{\rm TB}^{(j)}:=\left(\begin{array}{cc}
\frac{E_{\rm TB}-\epsilon_j}{V} & -1 \\
1 & 0 \end{array} \right).
\EEq
Now, assume that we have a circular lattice with infinite number of sites, indexed by $j$ where $j \in \{\ldots, -3,-2,-1,0,+1,+2,+3,\ldots\}$. Also, assume that $\epsilon_j$ being a random variable chosen from a uniform distribution. Note that, if $\psi_0$ and $\psi_1$ are known, then using this transfer-matrix approach we can iteratively determine (along both directions from origin) the probability amplitudes of an eigenstate as,
\BEqA
\left(\begin{array}{c}
\psi_{k+1} \\
\psi_{k} \end{array} \right)&=&M_{k}^{(+)}\left(\begin{array}{c}
\psi_{1} \\
\psi_{0} \end{array} \right) \;\; {\rm and} \nonumber \\ \nonumber \\
\left(\begin{array}{c}
\psi_{-k} \\
\psi_{-k-1} \end{array} \right)&=&M_{k}^{(-)}\left(\begin{array}{c}
\psi_{1} \\
\psi_{0} \end{array} \right),
\EEqA
where,
\BEqA
M_{k}^{(+)}&:=&\prod_{j=k}^{1}T_{\rm TB}^{(j)} \;\; {\rm and} \nonumber \\
M_{k}^{(-)}&:=&\prod_{j=-k}^{0}(T_{\rm TB}^{(j)})^{-1}\;.
\EEqA
Also, note that with $\epsilon_j$ being a random variable, the transfer-matrices $T_{\rm TB}^{(j)}$ are random symplectic matrices (so are $(T_{\rm TB}^{(j)})^{-1}$, as symplectic matrices form a group) having unit determinants. At this point, we invoke the tools of random matrix theory to show that the eigenstates in such a 1D disordered lattice are localized \cite{Thouless197493,prm1993Crisanti}. We specifically use F\"urstenberg theorem on product of random matrices, which states that, if $\{X_j\}$ is a set of uniformly distributed random matrices, then the limit,
\BEq
\lambda_{\rm 1}:=\lim_{k\to\infty}\frac{1}{k}\ln\left\|\prod_{j=1}^{k}X_j\right\|
\EEq
exists and $\lambda_{\rm 1}$ is usually referred to as \emph{maximum Lyapunov characteristic exponent}. The symbol $\|\ldots\|$ denotes the so-called operator norm of a matrix. F\"urstenberg also showed that $\lambda_{\rm 1}$ is nonrandom in general and if the random matrices are uniformly distributed and the determinant of each random matrix $X_{j}$ is unity then $\lambda_{\rm 1} > 0$. If we apply F\"urstenberg theorem for our case, it essentially means (remember $\psi_{j}$ is real for all $j$),
\BEq
|\psi_{k}| \sim e^{\lambda_{\rm 1}|k|}|\psi_{0}|, \;\; {\rm and} \;\;
|\psi_{-k}| \sim e^{\lambda_{\rm 1}|k|}|\psi_{0}|.
\EEq
Now, if we assume our lattice to be a circular one having many sites, then starting from origin the probability amplitudes increase exponentially on both directions and the exponential growth rate is the maximum Lyapunov characteristic exponent $\lambda_{\rm 1}$. For a closed lattice geometry, however, there is no guarantee that,
\BEq
\lim_{|k|\to\infty}|\psi_{k}|=|\psi_{-k}|
\label{eq:boundaryMatching}
\EEq
necessarily, and a mismatch in the closed boundary apparently seems paradoxical. We should emphasize at this point that although Eq. (\ref{eq:boundaryMatching}) cannot be satisfied in general, but it's sufficient for our purpose if it gets satisfied when $E_{\rm TB}$ is an eigenvalue of the Hamiltonian (\ref{eq:TBHamiltonian}), as opposed to any arbitrary real number. In fact, it has been observed that the probability amplitudes do match in the boundary for such a choice, and the inverse of the Lyapunov exponent $\lambda_{\rm 1}$ evaluated at any arbitrary energy tends to the \emph{localization length} $\xi_{\rm 0}$ (described below) of the eigenstate as $E_{\rm TB}$ gets closer to the eigenvalue of that eigenstate. This is known as Borland conjecture \cite{Borland1963}. Note that, we can choose any arbitrary site on a circular lattice as our origin and the exponential increase of probability amplitudes on both directions essentially indicates that on such a disordered lattice all the eigenstates are localized, which is consistent with the scaling theory of Anderson localization for one dimension \cite{0022-3719-6-10-009}.

\subsubsection{Measures of localization}
We define an eigenstate to be localized at the origin, if the probability amplitudes decrease exponentially with distance from the origin as,
\BEq
|\psi_{k}| \sim |\psi_{0}|e^{-|k|/\xi_{\rm 0}},
\EEq
where $\xi_{\rm 0}$ is referred to as the \emph{localization length} and can be defined in terms of the limiting probability amplitudes as \cite{prm1993Crisanti},
\BEq
\xi_{\rm 0}:=-\left[\lim_{|k|\to\infty}\frac{1}{|k|}\langle\ln{|\psi_k|}\rangle\right]^{-1},
\EEq
where the average (denoted by $\langle\ldots\rangle$) is taken over different configurations of lattice disorders. Notice that, for a uniformly extended state $\xi_{\rm 0}$ diverges, while it decreases for localized states and tends to zero for a Kronecker-$\delta$ like distribution on a discrete lattice.

While localization length is a measure of localization, it is not a unique one. Bell and Dean introduced another measure, called \emph{participation ratio} (denoted by ${\cal P}$ in this work), which also distinguishes between extended and localized states \cite{DF9705000055}. For a discrete lattice the participation ratio is defined as,
\BEq
{\cal P}:=\left[\sum_{j}|\psi_j|^4\right]^{-1},
\EEq
$\{\psi_j\}$ being the set of normalized probability amplitudes. Note that, on a discrete lattice with $N$ sites ${\cal P}=N$ for a perfectly extended state, and it tends to $+\infty$ as the number of sites increases. For a perfectly localized state ${\cal P}$ becomes unity.

Localized states can also be characterized by computing the moment for the position of the walker from its most likely location (assumed to be the origin). This approach is adopted by Yin et al. \cite{PhysRevA.77.022302}, where the second moment is used as a measure of localization. In our analysis in this work, we follow this moment-measure to quantify the localization, but instead of computing the second moment we compute the first moment, which is defined as,
\BEq
\mu^{(1)}:=\sum_{j}|j|\left(|\psi_{j}|^{2}+|\psi_{-j}|^{2}\right),
\label{eq:mu1}
\EEq
where $|j|$ denotes the distance from the origin and $(|\psi_{j}|^{2}+|\psi_{-j}|^{2})$ is the probability to find the particle at that distance from the origin. As far as characterizing the localization around a single lattice site is concerned, a case that is primarily considered in this work, this first moment essentially gives an effective length scale (from origin) in which the trajectory of the walker is restricted. Note that, $\mu^{(1)}$ can vary between $0$ and $N/2$, $N$ being the total number of lattice sites. For a localized QW we expect $\mu^{(1)}$ to be small in comparison to $N/2$ and the exact value denotes the expected range of its trajectory, which is more intuitive for visualizing the localization on a lattice.

\subsubsection{Localization via quantum walk}
The correspondence between Anderson localization and random walk was first established by Allen \cite{0022-3719-13-25-001} (also see Ref.~\cite{Thouless1980PR} for a brief summary), where it was shown using energy-time uncertainty relation that a random walk on a disordered 1D lattice eventually gets localized for any arbitrary disorder strength. The recent experimental realizations of localization of wavefunction in optical lattices via quantum walk \cite{Crespi2013,PhysRevLett.106.180403}, in fact, corroborate such an analogue. In this section, we, however, do not review the results obtained by Allen, but rather describe briefly how to achieve such a localization via DTQW. 

In order to introduce random diagonal disorders in the lattice, we perform a random phase rotation in the spin space of the walker after each Hadamard coin-tossing, which is given by,
\BEq
\hat{R}(\varphi_{k\downarrow})\ket{k,\downarrow}=e^{i\varphi_{k\downarrow}}\ket{k,\downarrow},
\hat{R}(\varphi_{k\uparrow})\ket{k,\uparrow}=e^{i\varphi_{k\uparrow}}\ket{k,\uparrow}.
\EEq
The rotation angles $\varphi_{k\downarrow}$ and $\varphi_{k\uparrow}$ are time-independent, but depends on the site as well as the spin-state of the walker. These angles are chosen at random from a uniform distribution between $-W\pi$ and $W\pi$, where $W$ is referred to as the disorder strength. In order to implement such random rotations within our scheme, we essentially need to perform a $\sigma^{z}$ rotation on each qubit in the lattice after every {\sf cross-Hadamard} gate, where the rotation angles are drawn from a uniform random distribution as mentioned above. For superconducting qubits such $\sigma^{z}$ rotations are in fact performed with random excursions of the qubit frequencies that closely mimics Anderson's original model where random energies are assigned to each lattice site. As shown below, it is possible to observe a transition from delocalized to localized states with increasing disorder strength under this scheme even for finitely many lattice sites.

In this context, we emphasize that there exists an unsettled controversy if the terminology--Anderson localization--should be used for localized quantum walks on a finite disordered lattice where the initial wavefunction of the walker is also localized \cite{Konno2010QIP,Wojcik2012PRA,PhysRevA.76.012315,PhysRevA.87.012314,Konno2013QIP,joye2010JSP}. We demonstrate here that if the walker is prepared initially on a particular lattice site, then it gets localized around its initial location if we perform QW in presence of a random static disorder on each site. We also investigate the case, where the walker is prepared as a uniform superposition of being in two diametrically opposite lattice sites at $t=0$, and observe that the final wavefunction of the walker gets localized around the initially populated sites when disorder is turned on. We, however, refer to this phenomenon as Anderson localization in this work (as also done in Ref.~\cite{Crespi2013} and \cite{PhysRevLett.106.180403}).

\section{Simulating DTQW with superconducting qubits}
\label{sec:QRWscq}
In this section, we first discuss the quantum walk with superconducting qubits in absence of any disorder, then demonstrate how it gets localized with increasing disorder strength, and finally consider a realistic model of $8$ coupled superconducting qubits, where the signature of such localized states can be observed. The results in this section are obtained via simulating the quantum circuit shown in Fig.~\ref{fig:generalCircuit} and assuming that the walker is initially localized in $\ket{0,\uparrow}$ state. However, at the end of Sec.~\ref{sec:QWwithDisorder}, we also consider a case where the walker initially occupies two diametrically opposite sites (in a superposition) instead of being on a single site.

\subsection{DTQW without disorder}
\label{sec:QRWscqNoDisorder}

\begin{figure*}[htb]
\centering
\subfloat[]{\includegraphics[angle=0,width=1.03\linewidth]{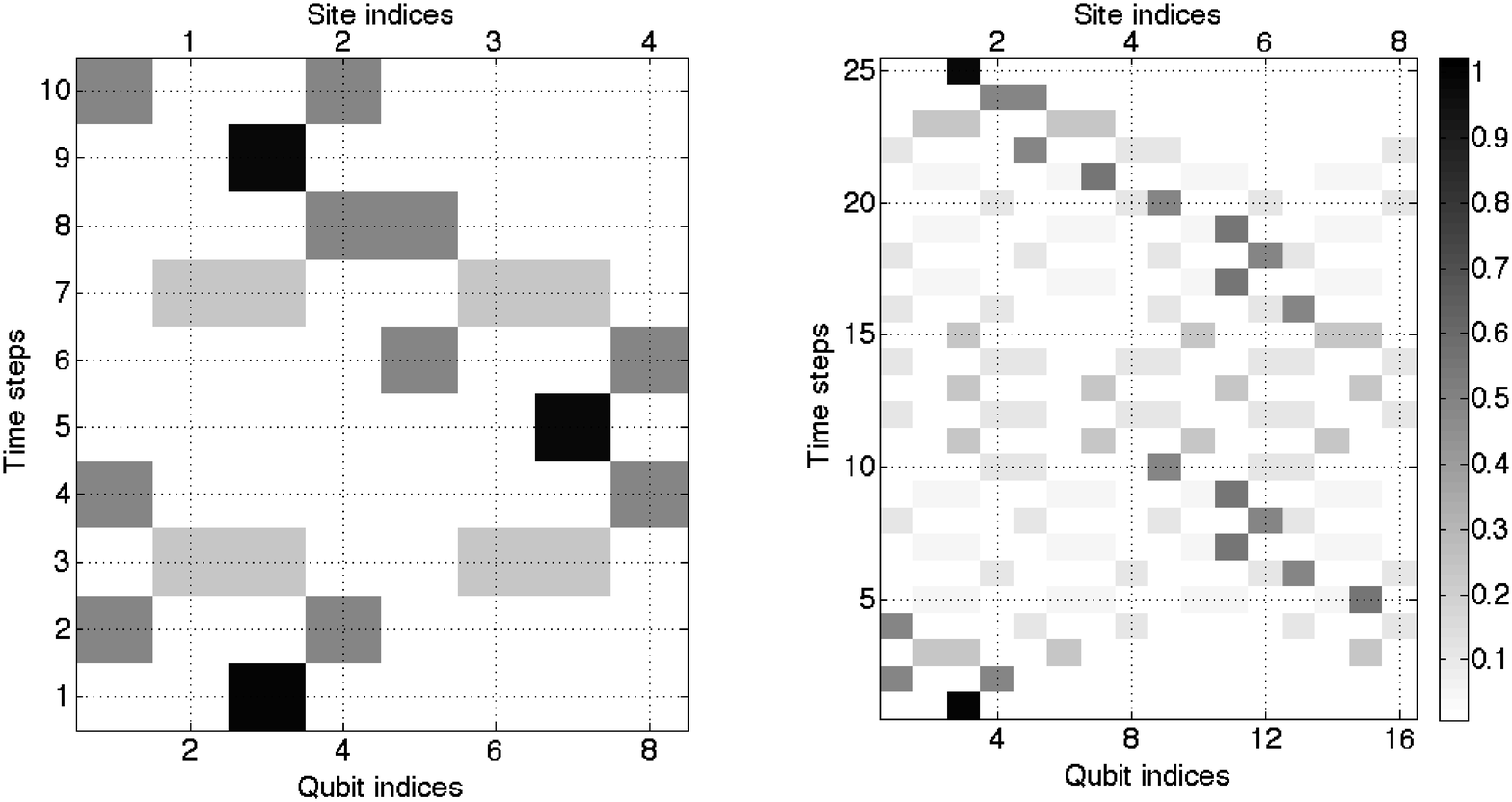}} \\
\subfloat[]{\includegraphics[angle=0,width=1.1\linewidth]{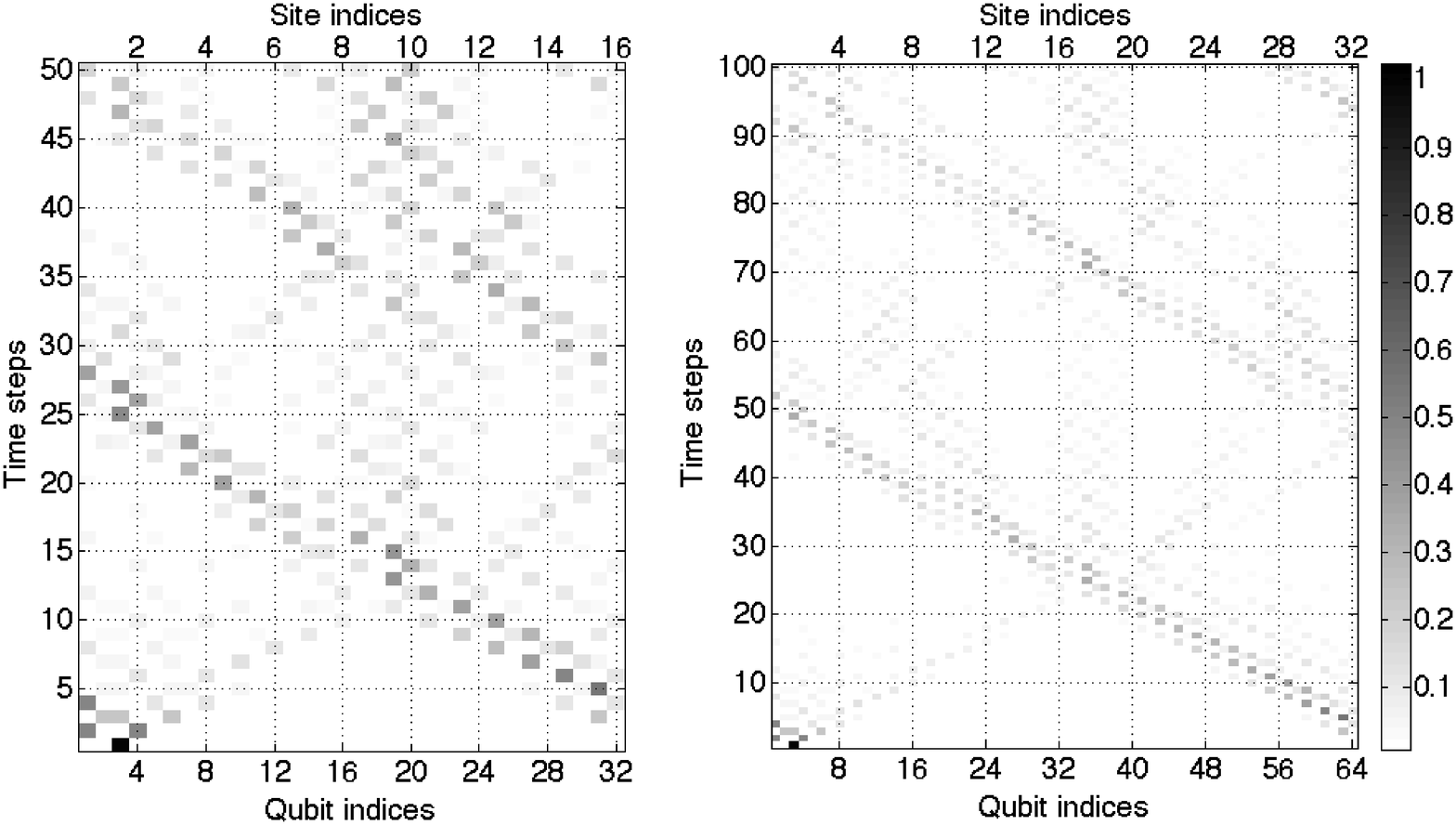}}
\caption{Simulation of quantum walk circuit (without disorder) in Fig.~\ref{fig:generalCircuit} for (a) $4$ and $8$ lattice sites and (b) $16$ and $32$ lattice sites. The horizontal axis denotes the qubit indices in the same order from top to bottom as in Fig.~\ref{fig:generalCircuit}, and the vertical axis denotes time. A periodicity in the probability distribution is observed in case (a), while no complete recurrence is found in case (b), which is consistent with earlier works \cite{chandrasekhar2010}.}
\label{fig:QRWsitesNoDisorder}
\end{figure*}

\begin{figure}[htb]
\includegraphics[angle=0,width=\linewidth]{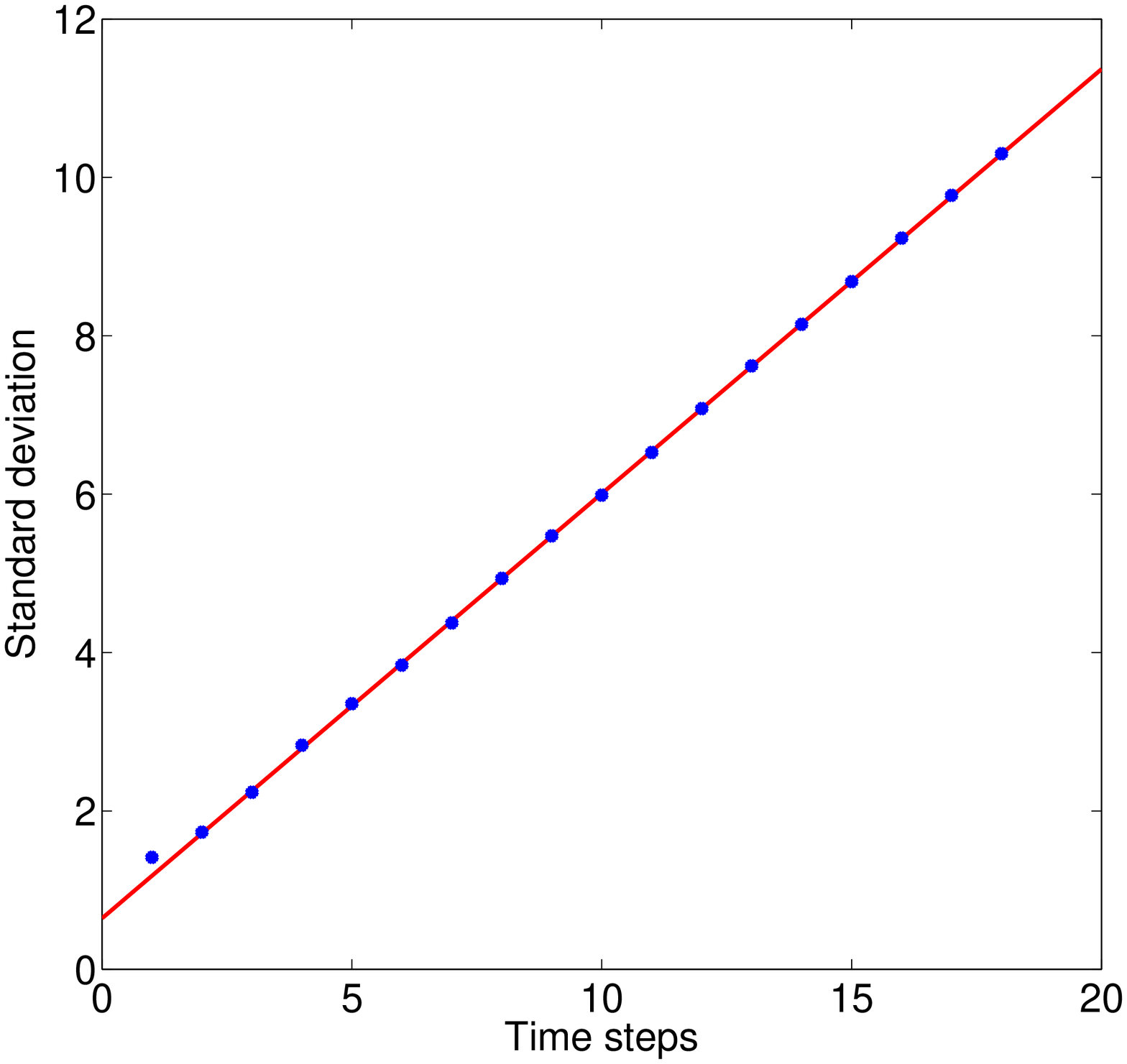}
\caption{(Color online) Plot showing the standard deviation of the probability distribution vs. time for $\Delta\theta=\pi/16$ on a circular lattice before the population reaches the boundary. The dots indicated the numerically evaluated points and the solid line denotes the best linear fit. Such a linear dependence denotes the ballistic spread of wavefunction as opposed to classical diffusion and considered to be a characteristic signature of quantum walk.}
\label{fig:sigmaPlot}
\end{figure}

In absence of any disorder, the circuit diagram shown in Fig.~\ref{fig:generalCircuit} simulates pure quantum walk on a circular lattice. We simulate our quantum circuit (Fig.\ref{fig:generalCircuit}) for $\Delta\theta=\pi/2$ ($4$ sites), $\pi/4$ ($8$ sites), $\pi/8$ ($16$ sites) and $\pi/16$ ($32$ sites). The results of our simulation are shown in Fig.~\ref{fig:QRWsitesNoDisorder}, where the qubit indices follow the same order from top to down as given by the quantum circuit in Fig.~\ref{fig:generalCircuit}. Since we consider the circular lattice, it is important to remember that the leftmost qubit is a nearest-neighbor to the rightmost qubit. The quantum circuit is simulated for many consecutive timesteps and each gray value in Fig.~\ref{fig:QRWsitesNoDisorder} denotes the probability to find the walker in that specific qubit.

The simulation in Fig.~\ref{fig:QRWsitesNoDisorder}(a) shows that when the total number of lattice sites is $4$ and $8$, the quantum walk repeats itself after every $7$ and $23$ timesteps respectively, which is a characteristic signature of DTQW on a circle as discussed in Sec.~\ref{sec:DTQRWonCircle}. However, for more than $8$ sites, such a complete recurrence becomes extinct and only fractional recurrence can be observed. In Fig.~\ref{fig:QRWsitesNoDisorder}(b), we have shown simulations with $16$ and $32$ lattice sites for $50$ and $100$ timesteps respectively, but no complete recurrence has been observed, which is consistent with earlier studies on DTQW on a circle \cite{chandrasekhar2010}.

As mentioned before, there is no difference between the signatures of a QW on a line or on a circle, until the population hits the boundary. In order to confirm that our quantum circuit in Fig.~\ref{fig:generalCircuit}, in fact, simulates a quantum walk, in Fig.~\ref{fig:sigmaPlot} we plot the standard deviation of the probability distribution (in the same unit of timesteps) for $\Delta\theta=\pi/16$. The plot shows standard deviations before population touches the boundary. A linear dependence is observed between the standard deviation and the time in that regime with a slope $\approx 3/5$, as found in earlier works \cite{PhysRevA.65.032310}. This ballistic spread of wavefunction as opposed to classical diffusion is a typical signature of quantum walk and almost independent of the initial state of the walker.

\subsection{DTQW with disorder}
\label{sec:QWwithDisorder}

\begin{figure}[htb]
\centering
\subfloat[]{\includegraphics[angle=0,width=\linewidth]{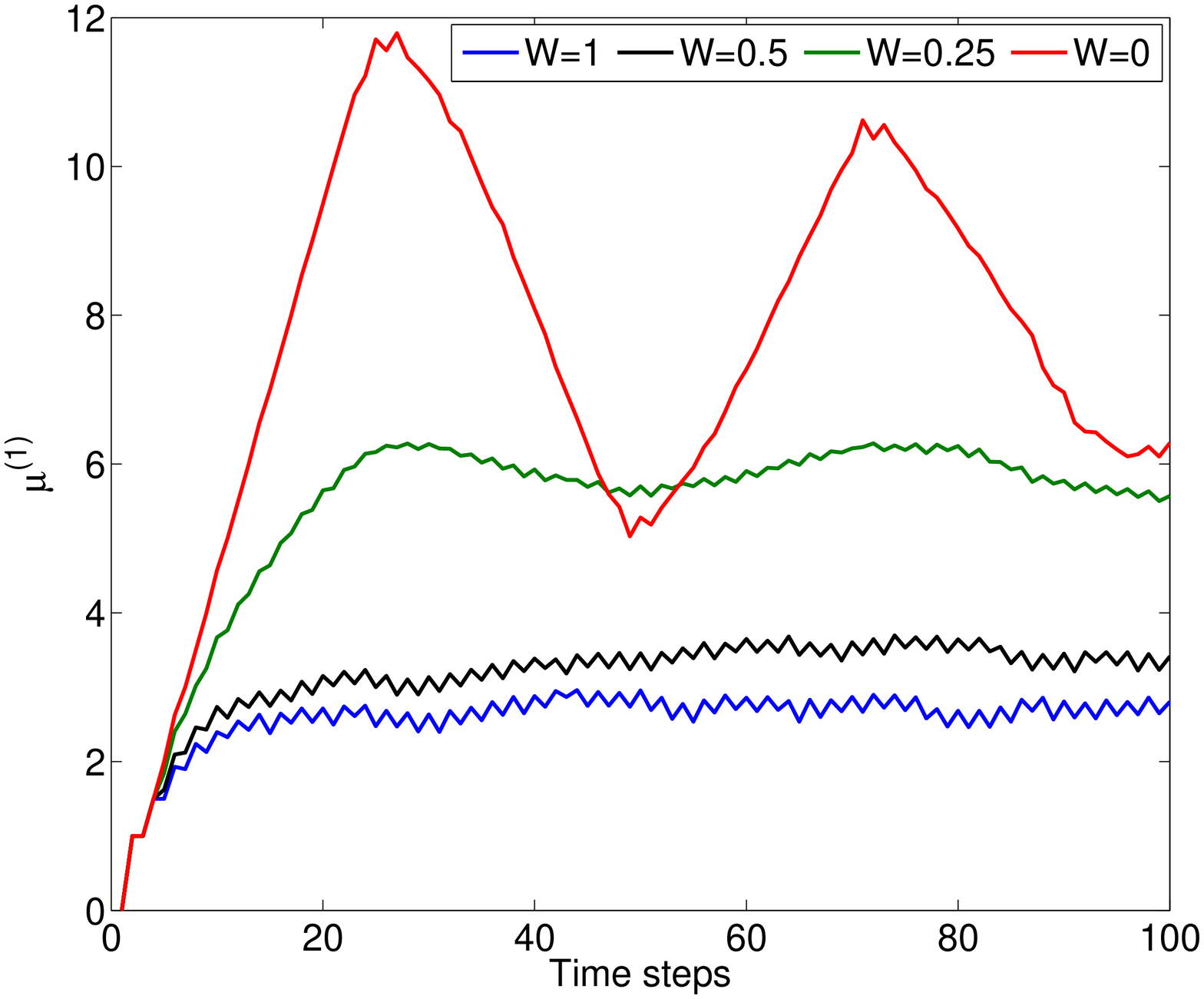}}\\
\subfloat[]{\includegraphics[angle=0,width=\linewidth]{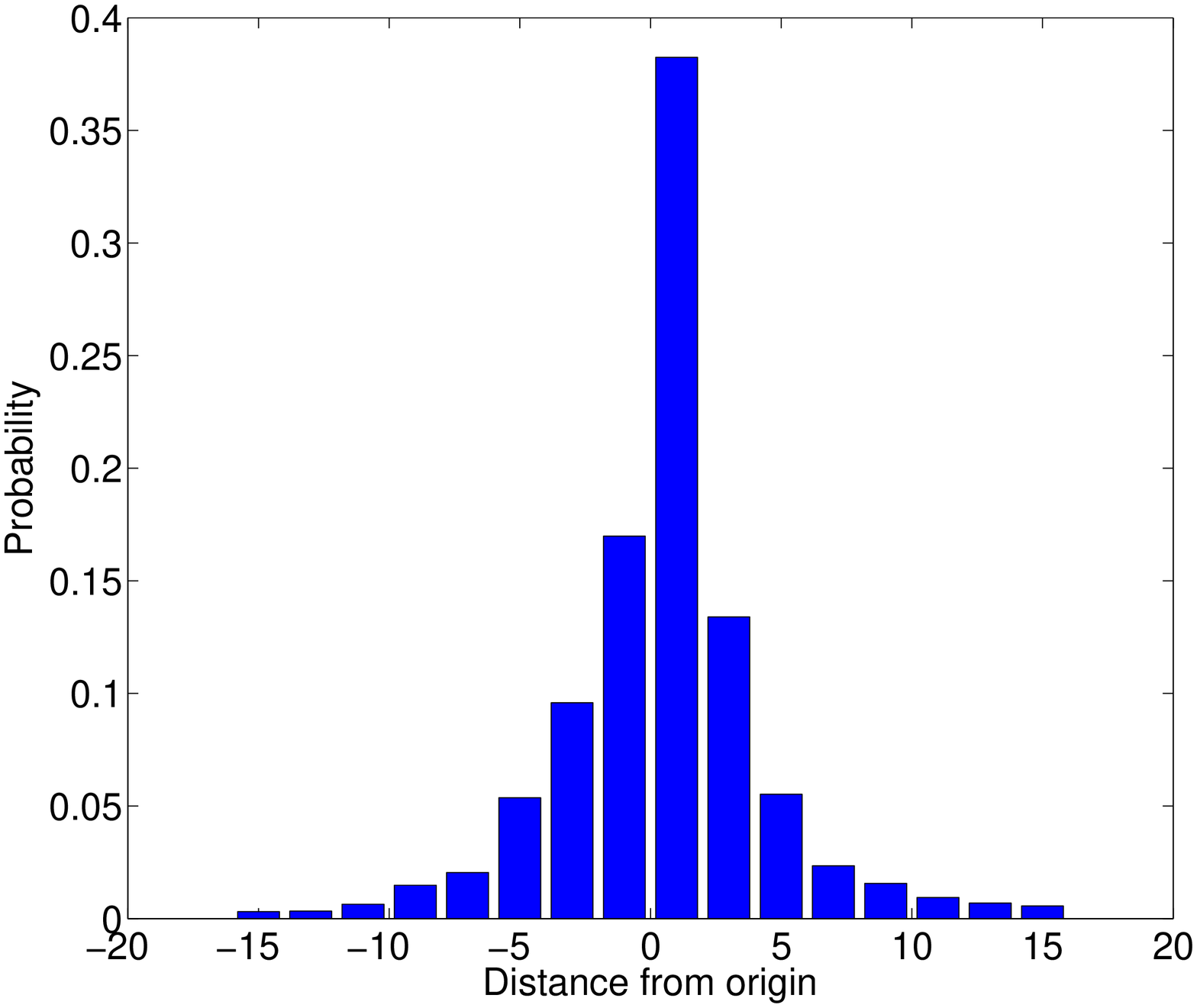}}\\
\caption{(Color online) Simulation of quantum circuit in Fig.~\ref{fig:generalCircuit} in presence of disorder for $\Delta\theta=\pi/16$. (a) The first moment $\mu^{(1)}$ is plotted against time for various disorder strengths $W$. A distinct signature of transition from the ballistic spread regime ($W=0$ case shown by uppermost solid red curve) to the localized walk ($W=1$ case shown by lowermost solid blue curve) is observed with increasing $W$. (b) Plot showing the probability distribution as a function of the distance from origin for $W=1$ case after $100$ timesteps. An exponential decay is observed in the probability of finding the particle away from the origin, which is a characteristic signature of wavefunction localization under disorder.}
\label{fig:QRWsitesWithDisorder}
\end{figure}

As discussed in Sec.~\ref{sec:DTQRWwithDisorder}, in a 1D lattice quantum walks get localized in presence of random static disorders. In the limit of an infinite lattice such a localization was predicted by Anderson, and 
the terminology `Anderson localization' gets used even in the context of localization on a finite lattice with an initially localized walker \cite{Crespi2013,PhysRevLett.106.180403}. In this section, we explore DTQW on a lattice of finitely many superconducting qubits. In this section, we concentrate entirely on the $\Delta\theta=\pi/16$ case (32 sites, 64 qubits), and demonstrate numerically how the quantum walk gets localized with increasing disorder strength.

As previously mentioned, we introduce disorders by performing some arbitrary $\sigma^{z}$ rotations on each qubit at each step in between the {\sf cross-Hadamard} and {\sf SWAP} gates. The rotation angles are chosen randomly from the interval $[-W\pi,W\pi]$, $W$ ($0 \leq W \leq 1$) denoting the strength of disorder. In order to characterize the localization of the DTQW, we compute the first moment $\mu^{(1)}$ (defined by Eq.(\ref{eq:mu1})), which essentially measures the expected absolute distance of the walker from the origin. Fig.~\ref{fig:QRWsitesWithDisorder} shows the results of our simulation for $\Delta\theta=\pi/16$, where each graph is computed by averaging over many possible realizations of disorders. In Fig.\ref{fig:QRWsitesWithDisorder}(a), we plot $\mu^{(1)}$ as a function of time, and observe a distinct signature of transition from the ballistic spread of the wavefunction to localized modes with increasing disorder strength. Note that, for $W=0$ the expected position of the walker varies almost (but not exactly as there is no complete recurrence in that regime) up to the farthest possible distance from the origin, and for $W=1$ it gets localized in the neighborhood of its initial position. Fig.\ref{fig:QRWsitesWithDisorder}(b) shows the probability distribution for $W=1$ case as a function of the distance from the origin (positive for clockwise, negative for anticlockwise) after $100$ timesteps and averaged over many realizations of disorders. A prominent exponential decay of the probability distribution is observed away from the origin, which is considered to be a typical signature of localization. 

\begin{figure}[htb]
\includegraphics[angle=0,width=\linewidth]{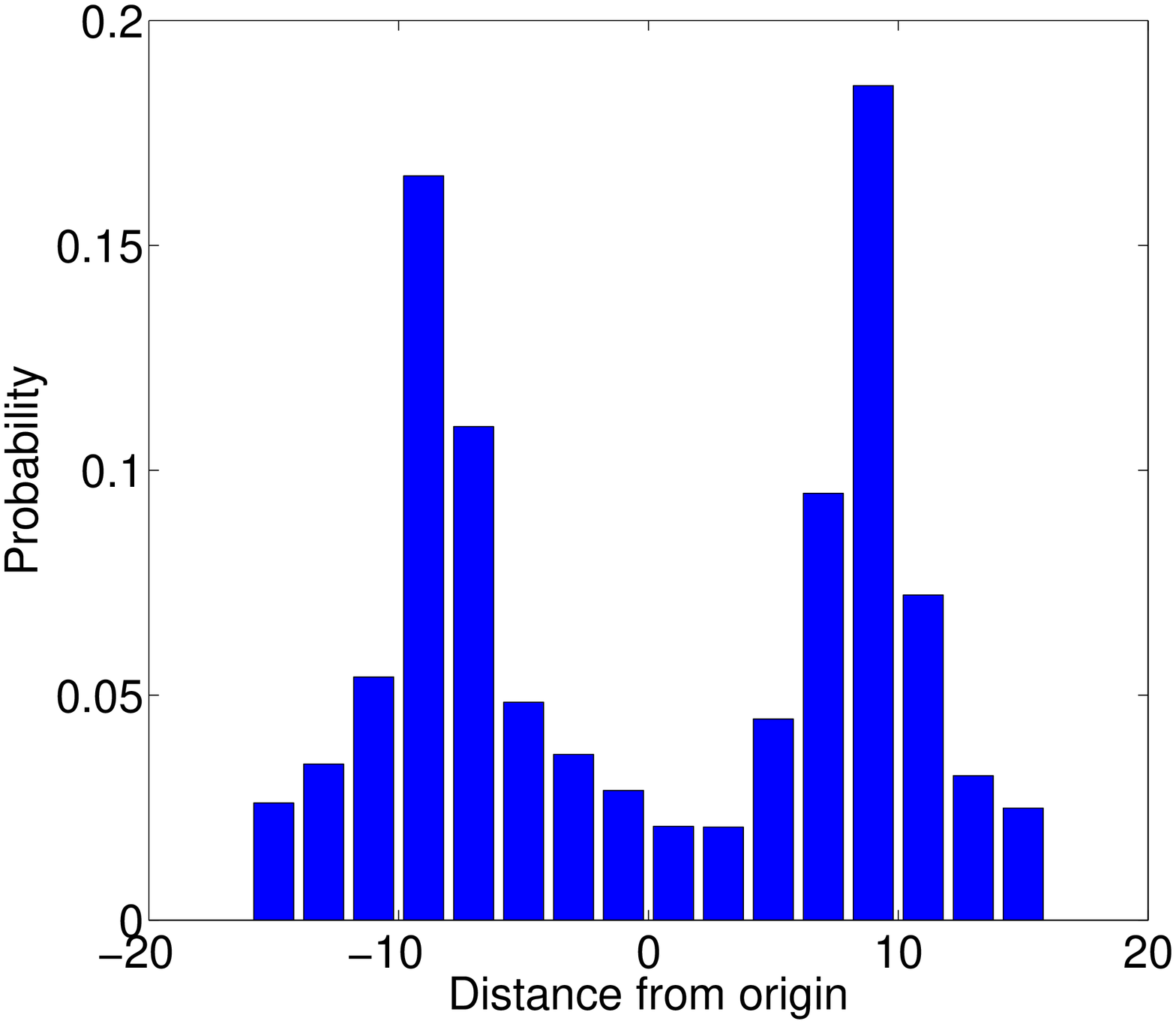}
\caption{(Color online) Plot showing the probability to find the walker on various sites on a circular lattice for W = 1 case after $100$ timesteps. Initially the walker is prepared in the state $(\ket{\pi/2,\downarrow}+\ket{-\pi/2,\uparrow})/\sqrt{2}$, as opposed to a single site. A bimodal distribution is observed denoting localization around each initially populated location.}
\label{fig:doublelocalization32sites50runs}
\end{figure}

We also investigate the case where the walker is initially prepared in two diametrically opposite sites on the circle instead of a single lattice site. Fig.~\ref{fig:doublelocalization32sites50runs} shows the probability distribution for $W=1$ after $100$ timesteps and averaged over many runs with different random sets of disorders for this case. We observe a bimodal distribution with peaks around the initially populated lattice sites, which indicates the localization of the wavefunction of the walker around its initial possible locations under QW in presence of random disorders on each lattice sites.

\subsection{Designing required gate operations}
In this section, we outline how to implement the required {\sf SWAP} and {\sf cross-Hadamard} gates with two coupled superconducting qubits. Note that our protocol only requires gate operations between nearest-neighbor qubits. Since we are assuming an architecture where the superconducting qubits are coupled with tunable couplers, for the purpose of gate design we only consider a two-qubit Hamiltonian, where other qubits are assumed to be decoupled from the system.

In a rotating frame, the Hamiltonian of two tunably coupled superconducting qubits is given by (in terms of Pauli matrices) \cite{PhysRevB.82.104522},
\BEq
H(t)=\frac{\Omega_{1}(t)}{2}\sigma_{1}^{z}+\frac{\Omega_{2}(t)}{2}\sigma_{2}^{z}+\frac{g(t)}{2}(\sigma_{1}^{x}\sigma_{2}^{x}+\sigma_{1}^{y}\sigma_{2}^{y}),
\label{eq:2qubitHamiltonian}
\EEq
$\sigma^{x}$, $\sigma^{y}$, and $\sigma^{z}$ being Pauli spin matrices and the subscripts denote the qubit indices. Since we are designing gates in the rotating frame, the terms $\Omega_{1,2}$ denote the detunings of qubit frequencies from the frequency of the rotating frame and for superconducting qubits we assume, $0 \; {\rm GHz} \leq \Omega_{1,2} \leq 2 \; {\rm GHz}$ and $-50\;{\rm MHz}\leq g \leq 50\;{\rm MHz}$. We also assume,
\BEqA
\Omega_{1}(t=0)&=&0 \nonumber \\
\Omega_{2}(t=0)&=&0 \nonumber \\
g(t=0)&=&0,
\label{eq:hamiltonianParameters}
\EEqA
where all the quantities are expressed in GHz in Eq.(\ref{eq:hamiltonianParameters}). Usually, superconducting qubits contain higher energy levels that are not considered in our Hamiltonian. While the auxiliary energy levels often play a crucial role in introducing leakage errors in an algorithm, for single-particle quantum walk with our approach their effect is negligible, as the entire system contains only one excitation. From the Hamiltonian (\ref{eq:2qubitHamiltonian}), it can be observed readily that any arbitrary $\sigma^{z}$ rotation can be performed on either qubit simply by qubit frequency excursions from the reference frequency, with the coupling turned off. In this section, we consider the two-qubit {\sf SWAP} and {\sf cross-Hadamard} gates. In order to perform these gate operations, we assume $\Omega_{\rm 1}=\Omega_{\rm 2}=0$ throughout and the coupling $g$ is varied with time.

\subsubsection{{\sf \it SWAP} gate}
\begin{figure}[htb]
\includegraphics[angle=0,width=\linewidth]{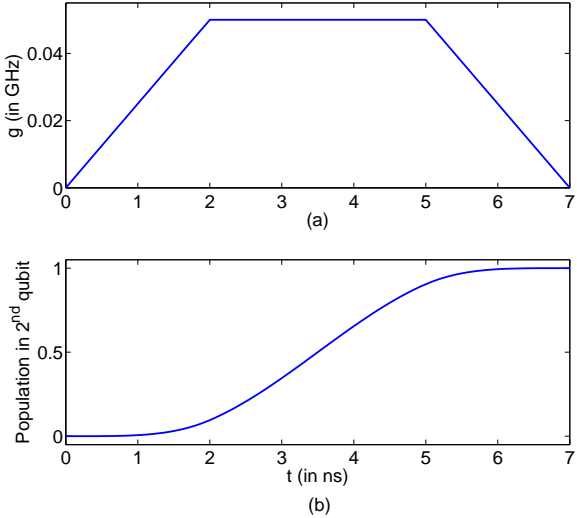}
\caption{(Color online) (a) A trapezoidal control pulse for the variable coupling strength $g$ that satisfies (\ref{eq:SWAPcondition}). (b) The probability to find the excitation on the $2^{nd}$ qubit is shown during the pulse.}
\label{fig:SWAPpulse}
\end{figure}
First, we discuss the pulse profile for the {\sf SWAP} gate. Two-qubit {\sf SWAP} gate (defined by Eq.(\ref{eq:SWAP})) acts as a $\sigma^{x}$ rotation in the single excitation subspace of the two qubits and acts as an Identity operation for other states. Note that, our two qubit Hamiltonian (\ref{eq:2qubitHamiltonian}), in the single excitation subspace (denoted by $H_{\rm s}$) can also be written as (with $\Omega_{\rm 1}=\Omega_{\rm 2}=0$),
\BEq
H_{\rm s}(t)=g(t)\left[\begin{array}{cc}
0 & 1 \\
1 & 0 \end{array} \right] \equiv g(t)\sigma^{x}.
\label{eq:Hses}
\EEq
In order to perform a $\pi$ rotation about $x$-axis, we need to choose a pulse for $g(t)$, such that,
\BEq
\int_{0}^{t_{\rm gate}}g(t)dt=\frac{\pi}{2}.
\label{eq:SWAPcondition}
\EEq
Fig.~\ref{fig:SWAPpulse}(a) shows a trapezoidal pulse that satisfies the constraint (\ref{eq:SWAPcondition}). We also showed how population gets transferred from the first qubit to the second with time under this pulse in Fig.~\ref{fig:SWAPpulse}(b). The parameters considered for this computation are consistent with the current superconducting control electronics and note that it is possible to perform {\sf SWAP} gate within $7$ nanoseconds. The {\sf SWAP} gate obtained under this pulse also contains a global phase in the single excitation subspace that can be nullified with post $\sigma^{z}$ rotations, a technique that is frequently used in superconducting quantum computing \cite{PhysRevA.87.022309}.

\subsubsection{{\sf \it cross-Hadamard} gate}
In the single excitation subspace, the {\sf cross-Hadamard} gate is given by,
\BEq
{\sf cross-Hadamard}_{\{\ket{01},\ket{10}\}}=\frac{1}{\sqrt{2}}\left[\begin{array}{cc}
1 & 1 \\
-1 & 1 \end{array} \right].
\EEq
We define the unitary rotation about any axis as,
\BEq
R_{\nu}(\phi)=e^{-i\frac{\phi}{2}\sigma^{\nu}},
\EEq
$\phi$ being the rotation angle and $\nu \in \{x,y,z\}$. It is easy to check that,
\BEq
{\sf cross-Hadamard}_{\{\ket{01},\ket{10}\}}=R_{y}\left(-\frac{\pi}{2}\right).
\EEq
Using the Euler angle decomposition, one can show,
\BEq
{\sf cross-Hadamard}_{\{\ket{01},\ket{10}\}} \equiv R_{z}\left(-\frac{\pi}{2}\right)R_{x}\left(\frac{\pi}{2}\right)R_{z}\left(\frac{\pi}{2}\right).
\EEq
The $\sigma^{z}$ rotations can be performed with qubit frequency excursions, as previously mentioned. We here discuss the $\pi/2$ rotation about $x$-axis. With the same reasoning as employed for {\sf SWAP} gate, we can derive a similar constraint for $R_{x}(\pi/2)$ as,
\BEq
\int_{0}^{t_{\rm gate}}g(t)dt=\frac{\pi}{4}.
\label{eq:cross-Hadamardcondition}
\EEq
\newline
Such a constraint can be satisfied with a similar trapezoidal pulse that encloses half the area enclosed by the {\sf SWAP} gate pulse in Fig.~\ref{fig:SWAPpulse}. According to (\ref{eq:cross-Hadamardcondition}), the {\sf cross-Hadamard} gate takes even less time than {\sf SWAP}. In order to discuss the effect of decoherence in section \ref{sec:decoherence}, we, however, modestly assume that each step in the circuit \ref{fig:generalCircuit} can be performed within $30$ ns.

\subsection{The $8$-qubit case}
\begin{figure}[htb]
\includegraphics[angle=0,width=\linewidth]{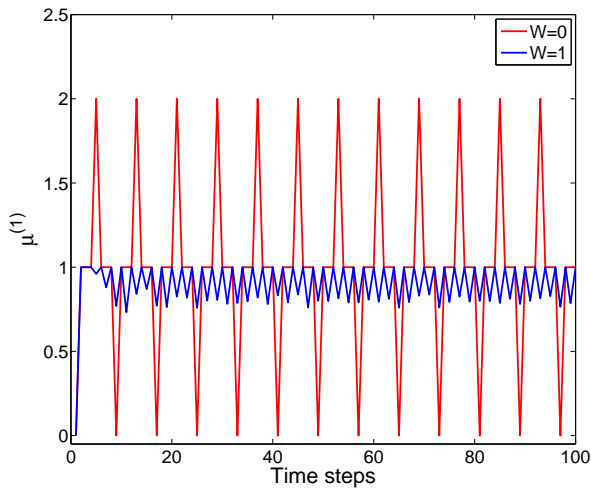}
\caption{(Color online) Simulation of quantum circuit in Fig.~\ref{fig:generalCircuit} in presence of disorder for $\Delta\theta=\pi/2$. The first moment $\mu^{(1)}$ is plotted against time for disorder strengths $W=0$ (dashed red) and $W=1$ (solid blue). A distinct signature of transition from the delocalized regime ($W=0$ case) to the localized regime ($W=1$ case) is observed.}
\label{fig:mu4sites300runs}
\end{figure}
So far, we outlined our implementation scheme for DTQW with superconducting qubits using our gate-based approach and observed that it is possible to simulate localized QW. Now, we concentrate on the $\Delta\theta=\pi/2$ case that involves $4$ lattice sites and requires $8$ qubits to realize. It is important to note that a circular lattice with $4$ sites offers an optimal architecture where for each site, there exists at least one other site (the diametrically opposite one) that is not the nearest neighbor of the previous one. Therefore, if we prepare the walker on a given site, it is interesting to explore if any signature of localization is observed where the probability to discover the walker in the neighborhood of its origin is maximum in comparison to finding it on the diametrically opposite position. This motivates us to cast a special attention to the $8$-qubit case.

Here we demonstrate numerically that the signature of two distinct regimes, the ballistic regime in absence of disorder and the localized QW regime in the disordered lattice, can be clearly distinguished even for such a few-qubit system. Fig.~\ref{fig:mu4sites300runs} shows the result of our simulation for this case, where the first moment $\mu^{(1)}$ is plotted against time for $W=0$ and $1$. Note that, while the walker travels through the entire lattice periodically in absence of disorder, it gets localized quite noticeably around the origin when disorder is turned on. Observing such a localized quantum walk for $\Delta\theta=\pi/2$ requires only $8$ nearest-neighbor-coupled qubits, an architecture that is already within reach of current superconducting qubit technology and could be realized in the near future.

\subsection{Effect of intrinsic errors and decoherence}
\label{sec:decoherence}
Superconducting qubits possess some additional states other than the computational $\ket{0}$ and $\ket{1}$ states. These higher energy levels play a significant role in producing intrinsic leakage errors in any quantum computation with such qubits, not only because they exist, but also these states often get utilized for designing some quantum gates \cite{PhysRevA.87.022309,Ghosh2013arXiv}. However, some recent advancements show that designing high-fidelity two-qubit entangling gates is possible by suppressing such leakage errors below $10^{-4}$ \cite{PhysRevA.87.022309}. In our simulation, such intrinsic errors are not considered because for single-particle quantum walk within our scheme, the entire lattice always remains in the single-excitation subspace, for which the effect of such higher-energy-level-induced errors remains negligible anyway.

Another challenge to perform any quantum computation with superconducting qubits is the decoherence. A tremendous progress has been made along this direction in the past few years and a superconducting qubit (called `Xmon') with $T_{\rm 1}\approx 44 \;{\mu}s$ has recently been reported \cite{PhysRevLett.111.080502}. Such a long coherence time can be achieved for these Xmon qubits without any 3D cavity, and therefore, they remain as one of the best candidates for fabricating long 1D or 2D lattices of coupled superconducting qubits. As we showed earlier, each timestep in our quantum circuit \ref{fig:generalCircuit} can be performed within $30$ ns., which essentially means that $100$ such timesteps would require only $3$ $\mu{s}$. Assuming that decoherence is only dependent on the total simulation time and affect each qubit individually, we argue that the effect of decoherence in realizing quantum walk or observing localization with our approach remains negligible since such effects become prominent within $100$ timesteps (as shown in Fig.~\ref{fig:QRWsitesWithDisorder} and \ref{fig:mu4sites300runs}) and the required simulation time ($3$ ${\mu}s$) is order of magnitude smaller than the energy relaxation time of the current superconducting qubits, which is $44$ $\mu$s.

\section{Conclusions}
\label{sec:conclusions}
The motivation to realize interesting quantum transport processes with superconducting qubits emerges from the fact that the superconducting qubits have long coherence times and high degree of scalability. However, the challenge to simulate such processes via quantum walk comes from their stationary nature, that superconducting qubits are fabricated with on-chip Josephson junctions and therefore, cannot hop from one lattice site to another. Our proposal in this work offers a solution to this problem. We treat the excitations of the superconducting qubits as our walkers and artificially introduce their spin degrees of freedom by placing two qubits (gray and black) at each lattice site and adopting the convention that the walker is in $\ket{\uparrow}$ state if the excitation is in the gray qubit, and $\ket{\downarrow}$ state if it is in the black qubit. We numerically demonstrated that such a mapping is capable of simulating DTQW on a one-dimensional lattice. A quantum circuit is discovered for this purpose and the gate-design protocols are discussed. As an additional benefit, it has also been shown that lattice disorders can be introduced and controlled for each individual site within this scheme. While it is possible to introduce both static and dynamic disorders in this protocol, we here concentrated on random diagonal static disorders and observed a transition of the wavefunction of the walker from its ballistic regime to the localized mode, a phenomena that was first predicted by Anderson for infinitely long lattices. An 8-qubit experiment is also proposed where the signature of localized quantum walk can be observed with increasing disorder strength. Conducting such a few-qubit experiment is already within reach of the current superconducting technology and could be implemented in near future. While in this work we primarily concentrated on single-particle QW, a generalization of this scheme for multi-particle QW is also possible and is discussed elsewhere \footnote{J. Ghosh et al., unpublished.}. Our proposal thus opens up the possibility to explore various quantum transport processes using promising superconducting qubits. A detailed investigation on multi-particle QW in presence of static and dynamic disorders could be a possible future direction of this research.

\begin{acknowledgments}
This research was funded by NSERC, AITF and University of Calgary's Eyes High Fellowship Program. I thank Barry Sanders for many illuminating comments and suggestions as well as his careful reading of the manuscript. I also gratefully acknowledge useful discussions with Guilherme M. A. Almeida, Andrew Childs, Sandeep Goyal, Miguel A. Martin-Delgado, Mario Mulansky, Emily Pritchett and Pedram Roushan.
\end{acknowledgments}

\bibliography{QRWwithDisorder}

\end{document}